\documentclass[sigplan,screen,nonacm]{acmart}

\usepackage[]{hyperref}

 \def\fig{Figure~}

\def\tab{Table~}

\definecolor{mygray}{gray}{0.9}

\usepackage{dsfont}

\usepackage{xspace}

\makeatletter
\def\onedot{\ifx\@let@token.\else.\null\fi\xspace}
\makeatother
\def\ie{\emph{i.e}\onedot,\xspace}

\def\ie{\emph{i.e}\onedot,\xspace} 
 
\def\etc{\emph{etc}\onedot} 
\def\name{CMSwitch~}
\newcommand\update[1]{\textcolor{black}{#1}}

\usepackage{multirow}
\usepackage{colortbl}
\usepackage{tabularx}
\usepackage{algorithm}
\usepackage{algorithmic}
\copyrightyear{2025}
\acmYear{2025}
\setcopyright{rightsretained}
\acmConference[ASPLOS '25]{Proceedings of the 30th ACM International Conference on Architectural Support for Programming Languages and Operating Systems, Volume 2}{March 30-April 3, 2025}{Rotterdam, Netherlands}
\acmBooktitle{Proceedings of the 30th ACM International Conference on Architectural Support for Programming Languages and Operating Systems, Volume 2 (ASPLOS '25), March 30-April 3, 2025, Rotterdam, Netherlands}
\acmDOI{10.1145/3676641.3716248}
\acmISBN{979-8-4007-1079-7/25/03}

\begin{document}

\title{Be CIM or Be Memory: A Dual-mode-aware DNN Compiler for CIM Accelerators}

\author{Shixin Zhao}
\affiliation{%
    \institution{Institute of Computing Technology, Chinese Academy of Sciences, University of Chinese Academy of Sciences}
  \city{Beijing}
  \country{China}
  \postcode{100190}
}
\email{zhaoshixin18@mails.ucas.ac.cn}

\author{Yuming Li}
\affiliation{%
  \institution{Institute of Computing Technology, Chinese Academy of Sciences, University of Chinese Academy of Sciences}
  \city{Beijing}
  \country{China}
}
\email{liyuming22@mails.ucas.ac.cn}

\author{Bing Li}
\affiliation{%
  \institution{Institute of Microelectronics, Chinese Academy of Sciences}
  \city{Beijing}
  \country{China}}
\email{libing2024@ime.ac.cn}

\author{Yintao He}
\affiliation{%
  \institution{Institute of Computing Technology, Chinese Academy of Sciences, University of Chinese Academy of Sciences}
  \city{Beijing}
  \country{China}}
\email{heyintao19z@ict.ac.cn}

\author{Mengdi Wang}
\affiliation{%
  \institution{State Key Lab of Processors, Institute of Computing Technology,  Chinese Academy of Sciences}
  \city{Beijing}
  \country{China}}
\email{wangmengdi@ict.ac.cn}

\author{Yinhe Han}
\affiliation{%
  \institution{State Key Lab of Processors, Institute of Computing Technology,  Chinese Academy of Sciences}
  \city{Beijing}
  \country{China}}
\email{yinhes@ict.ac.cn}

\author{Ying Wang}
\authornote{Corresponding author.}
\affiliation{%
  \institution{State Key Lab of Processors, Institute of Computing Technology,  Chinese Academy of Sciences}
  \city{Beijing}
  \country{China}}
\email{wangying2009@ict.ac.cn}

\renewcommand{\shortauthors}{Shixin Zhao et al.}

\begin{abstract}
Computing-in-memory (CIM) architectures demonstrate superior performance over traditional architectures. \update{To unleash the potential of CIM accelerators, many compilation methods have been proposed, }
focusing on application scheduling optimization specific to CIM. 
However, existing compilation methods often overlook CIM's capability to switch dynamically between compute and memory modes, which is crucial for accommodating the diverse memory and computational needs of real-world deep neural network architectures, especially the emerging large language models.
\update{To fill this gap, we introduce CMSwitch, a novel compiler to optimize resource allocation for CIM accelerators with adaptive mode-switching capabilities, thereby enhancing the performance of DNN applications.}
Specifically, our approach integrates the compute-memory mode switch into the CIM compilation optimization space by introducing a new hardware abstraction attribute. Then, we propose a novel compilation optimization pass that identifies the optimal network segment and the corresponding mode resource allocations 
using dynamic programming and mixed-integer programming. CMSwitch uses the tailored meta-operator to express the compilation result in a generalized manner.
Evaluation results demonstrate that CMSwitch achieves \update{an average speedup of 1.31$\times$ compared to existing SOTA CIM compilation works, highlighting CMSwitch's effectiveness in fully exploiting the potential of CIM processors for a wide range of real-world DNN applications.}
\end{abstract}



\maketitle 

\section{Introduction}
The Computing-In-Memory (CIM) architecture is highly regarded for enabling in-situ computation \cite{chen2020survey,ankit2019puma,biswas2018conv,eckert2018neural,song2017pipelayer,chi2016prime,shafiee2016isaac}. CIM minimizes frequent data transfers and enhances the parallel execution of matrix multiply-and-accumulate (MAC) operations, resulting in notable performance improvements. Compared to conventional architectures, CIM significantly mitigates persistent memory wall problem \cite{wulf1995hitting} and demonstrates strong competitiveness in data-intensive applications \update{especially deep neural network (DNN) inference} \cite{yang2020retransformer,shafiee2016isaac,sridharan2023x}.

\begin{figure}[t]
    \centering
    \includegraphics[width=0.9\linewidth]{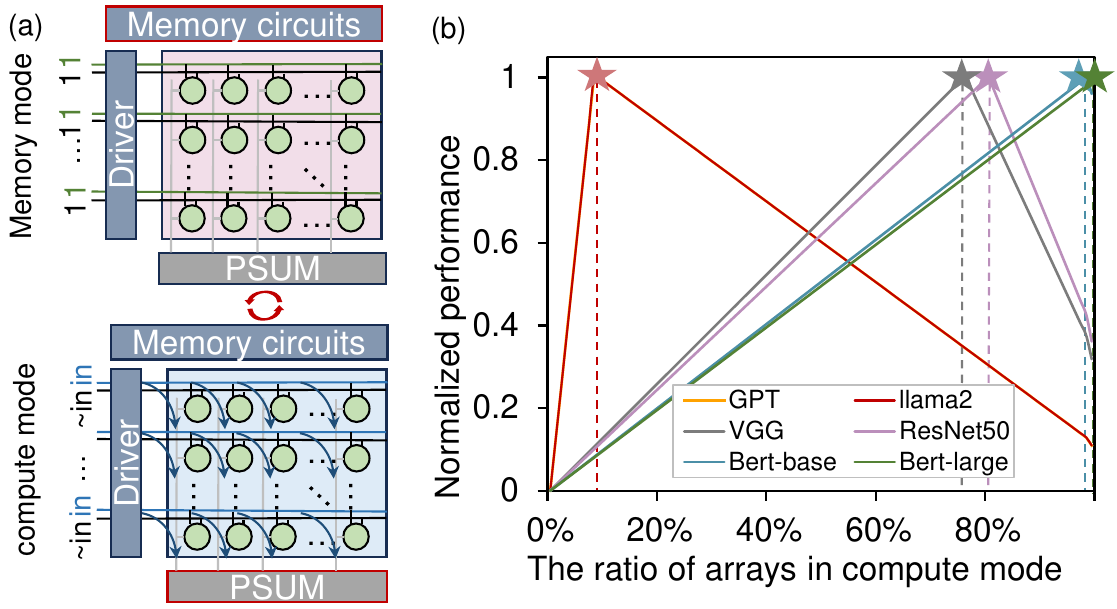}
    \caption{(a) CIM switching between memory and compute mode by setting up control signals to the input driver; (b) Normalized performance variation with the ratio of arrays in compute mode changes. Please note that putting more CIM arrays in compute mode deprives them of the chance of working as scratchpad memory for storing and loading intermediate data, e.g. activations. CIM arrays in compute mode must store static data, i.e. pre-determined weights.}
    \label{fig:intro}
\end{figure}

To enhance efficiency and fully realize the potential of the CIM accelerators, researchers have explored various compilation optimization techniques, aimed at various CIM architectures such as resistant RAM and SRAM-based solutions \cite{qu2024cim,sun2023pimcomp,farzaneh2023c4cam,han2021polyhedral,siemieniuk2021occ,drebes2020tc}. These compilation tools significantly reduce entry barriers for users adopting the CIM architecture and support the widespread deployment of CIM chips.
Earlier compilation tools for CIMs assume that neural network weights are pre-loaded into memory. These tools formulate an optimized policy for weight mapping and devise a computation scheduling scheme across various operational granularities to maximize memory resource utilization and enhance computational performance. \cite{qu2024cim,siemieniuk2021occ,ankit2019puma}.
Although there have been significant improvements in compilation optimization techniques for CIM architecture, current methods still consider the memory and compute resources on the chip static, which does not accurately represent the modern advancements in CIM designs.
In practice, many modern CIM designs feature dual-mode memory arrays that can dynamically switch between memory and compute modes~\cite{kim202316,yan20221,guo202328nm,yue202115}. As depicted in \fig\ref{fig:intro}(a), the CIM array transitions between these modes by resetting the input driver. This dynamic functionality broadens the compiler's optimization possibilities for CIM mapping, \update{enhancing DNN application performance}. Previous compiler-level optimization efforts did not fully exploit these opportunities, missing out on the benefits of dual-mode CIM arrays.


Moreover, different real-world DNN architectures have distinct memory and computation resource requirements. \fig\ref{fig:intro} (b) depicts the varying demands on the memory and computation resource of different DNN models (\ie CNNs \cite{he2016deep}, LLaMA \cite{touvron2023llama}, GPT~\cite{brown2020language}, \etc) to reach the optimal performance on the CIM chip.
Convolutional neural networks (CNNs) have relatively high arithmetic intensity (FLOPs/ Memory OP) and demand a higher ratio of compute to memory resources on CIM. For instance,  ResNet50 has an average arithmetic intensity of 66, and its performance reaches the highest point when the ratio of compute to memory resource reaches almost 80\%. Thus, some typical CNNs require more CIM arrays working in compute mode, when they already have sufficient CIM arrays configured as the on-chip scratchpad memory for activation caching.
 In contrast, Transformer-based models typically have much lower arithmetic intensity. For instance, the generative model LLaMA 2 has an average arithmetic intensity of around 2 for single batch inference and \fig\ref{fig:intro} (b) depicts that LLaMA 2 garners the best performance when the ratio of compute to memory in CIM arrays is about 10\%, which means it is better to offer more on-chip random memory for activations and KV cache rather than to increase compute-power, given that it is almost impossible to cache all the massive parameters of large language models on a single CIM chip. Although this conclusion drawn from Figure 1 only makes sense for certain models and hardware configurations, like the on-chip memory space, main memory bandwidth, etc, it reveals a fact that it is not necessarily correct to assume all CIM arrays should be put at the compute mode as in prior compilation works.
 Moreover, the requirements for memory and compute CIM arrays of the same model may vary across different layers or stages of execution. 
 Therefore, a compiler customized for dual-mode CIM that optimizes the memory and compute mode of the CIM array is significant.

\update{In this work, we propose a novel CIM compiler that takes the CIM mode switch into account and co-adjusts the CIM working mode and the mapping of the DNN applications in the context of dual-mode CIMs. 
Specifically, for a given target CIM architecture and the neural network workload, the proposed compiler can determine the arrays' mode being the compute or memory, and the optimal allocation of those arrays.} Once the mode-switch decision is made, the compiler also schedules operators on the respective arrays
to achieve optimal performance.

However, to achieve this goal, we have to address the following two challenges: 
(1) \textbf{Exponential space expansion}: 
In dual-mode CIM, each CIM array can work in memory or compute mode. \update{Consequently, the problems of array mode selection and weight mapping are entangled in the deployment of target DNNs}, which constitutes a larger exploration space for the compiler.
For instance, with $m$ arrays in CIM, there are $2^m$ choices of the mode allocation during the compilation. It is proved that model scheduling for CIM is already a complicated problem with the optimization space about polynomial complexity, dual-mode CIM will face a $2^m$ times larger space with an exponential level. 
Therefore, we have to formalize such a jointly optimized exploration space that combines the allocation and mapping decisions, along with a search strategy when designing our compiler.
(2) \textbf{Dual-mode switching schedule}: When scheduling each DNN operator in a dual-mode CIM, we should deliberately determine the mode of arrays as the number of arrays in different modes also affects the efficiency of the current operator and the scheduling of subsequent operators. Thus, the compiler for dual-mode CIM must account for the interdependence between array mode scheduling and weight mapping. The separate treatment of mapping and scheduling in previous compilers for CIM with fixed-mode arrays is insufficient for enhancing performance in dual-mode CIM architectures.
\update{Thus, we propose a holistic optimization framework that integrates DNN mapping and array mode scheduling for dual-mode CIMs.} 

To address these challenges, our approach at first provides the hardware abstraction of dual-mode CIM accelerators \update{based on CIM-MLC}~\cite{qu2024cim} so that the dimension of CIM reconfiguration can be fused into the original mapping/scheduling space of CIM as a formalized optimization space.
\update{Second, to make the joint optimization problem tractable for modern large-scale neural networks, we employ a divide-and-conquer two-step policy, co-optimizing the array mode switch, allocation, and mapping of neural networks.} Given that CIM memory space often cannot accommodate the entire model on the chip, the network must be executed in segmented partitions in serial. This is a common trend with billion-scale large language models.
\update{We first utilize dynamic programming (DP) to network segmentation. The overhead introduced by the array mode switch is taken into account when applying DP for global optimization.}
Afterward, we use mixed integer programming to automatically explore the optimization space for operator mapping with tunable hardware resources within each segment. The compiled results are then output in a meta-operator flow marked with memory-compute switch information. 

Specifically, the main contributions of this work include:
\begin{itemize}
    \item 
    \update{To support various DNNs including nowadays large language models,
    we introduce CMSwitch, a novel dual-mode-aware CIM compiler that leverages the mode switch capability of compute/memory of CIM arrays to meet diverse DNN application requirements. We formalized the joint-optimization problem of mode-switch, mapping, and scheduling for standalone CIM accelerators, and released the first compiler aware of this important CIM feature.}
    \item We comprehensively consider the challenges and opportunities brought by mode switching. Without causing too much exploration overhead, we propose a two-step optimization strategy to make the compilation process converge at the optimal design point of the large joint design space. 
    By formalizing the overhead and performance improvement introduced by CIM mode switch, we employ DP and MIP to determine the optimal network segment in temporal and allocate compute-memory resources in spatial.
    \item We evaluate the \name across a set of DNN benchmarks. \update{Compared with state-of-the-art compilation works \cite{qu2024cim}, \name achieves average inference speed improvement by 1.31$\times$.} We also verify \name for various workload scales, demonstrating robust dual-mode-aware compilation support for \update{diverse DNN architecture demands}. It is proved the proposed compiler shows especially great potential for popular large models that cannot be fitted into the on-chip memory.
\end{itemize}


\section{Background}
\subsection{\update{Computing-In-Memory Accelerator Architecture}}
\begin{figure}[t]
    \centering
    \includegraphics[width=\linewidth]{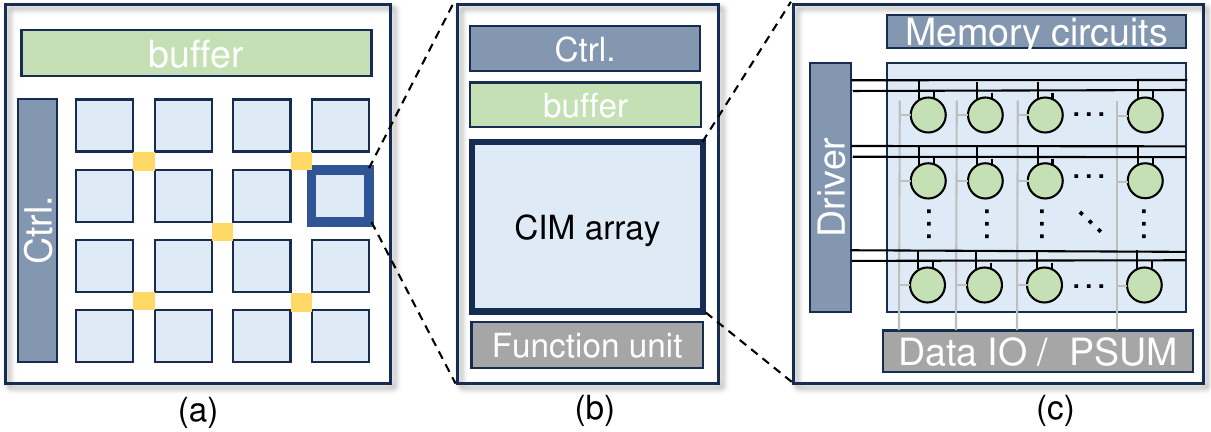}
    \caption{Hierarchical CIM architecture (a) CIM core with CIM arrays and the corresponding peripheral units; (b) CIM array and the corresponding peripheral units; (c) CIM array.}
    \label{fig:CIM_archi}
\end{figure}
\update{As depicted in \fig \ref{fig:CIM_archi} (a)-(c), independent CIM-based DNN accelerators are typically structured as hierarchical architecture comprising multiple CIM cores. Each core integrates a CIM array along with its peripheral buffer and circuitry. This design enables in-situ computation within the memory, thereby mitigating the data transfer bottlenecks commonly seen in conventional architectures that separate computation and memory. Prior researches \cite{le202364,wan2022compute,chen2020survey,ankit2019puma,eckert2018neural,biswas2018conv,seshadri2017ambit,song2017pipelayer,li2017drisa,shafiee2016isaac,chi2016prime,ahn2015pim,li2018reram,joardar2019regent,qu2020raqu} have proposed various CIM accelerators, providing robust support for high-performance computing and naturally aligning with large-scale parallel computing applications such as DNN inference. }

 
\subsubsection{Dual Modes CIM Array}

The dual-mode CIM array can operate as both a memory and compute unit when applying a slight enhancement on the input or output drivers~\cite{kim202316,yan20221,yue202115,chih202116,fujiwara20225,wang2022dimc,spetalnick202240nm,ankit2020panther}.

\begin{figure}[t]
    \centering
    \includegraphics[width=\linewidth]{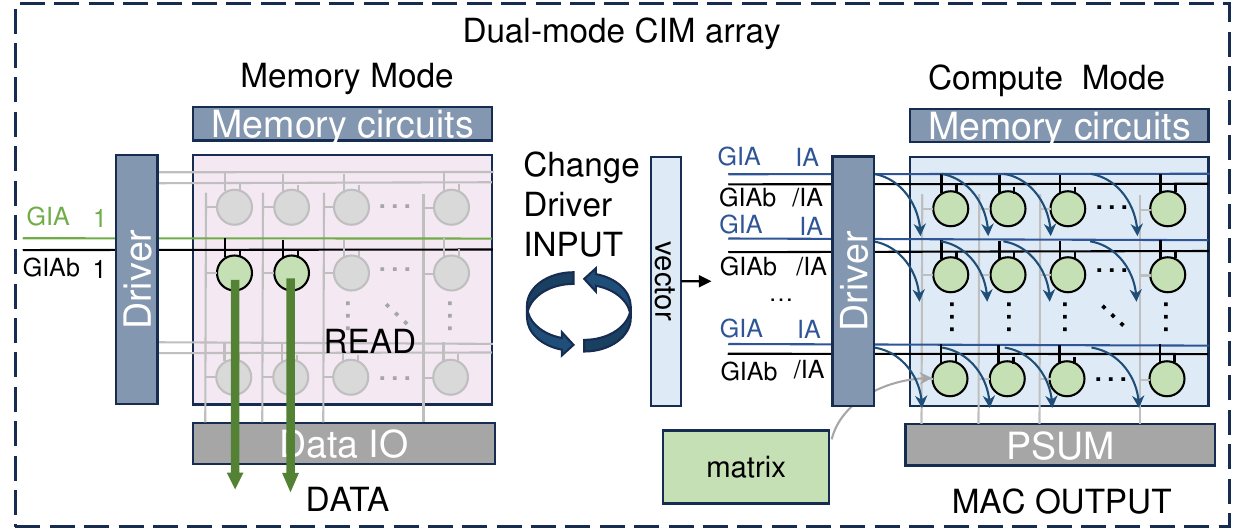}
    \caption{Dual-mode CIM Array.}
    \label{fig:cim_switch}
\end{figure}
As illustrated in \fig \ref{fig:cim_switch}, switching between memory and compute modes of CIM arrays can be achieved by altering the array's driver inputs, as demonstrated by DynaPlasia \cite{kim202316}.
This mode-switching functionality is controlled by modifying the input signals on the global lines.
When the Global Input Activation line (GIA) and its complement (GIAb) are set to a high state (1), the array functions in memory mode, allowing standard memory read-write operations. 
Conversely, when the GIA and GIAb are configured as input activation (IA) and inverse of input activation IA (/IA), respectively, the array operates in compute mode, performing bit-series multiplication-addition operations.  

\subsubsection{CIM Compute Paradigm}
When CIM arrays perform computations, they enable the multiply-accumulate (MAC) operations to be executed entirely within the array in parallel, as illustrated in the \fig \ref{fig:cim_switch} (right). This architecture inherently supports matrix-vector multiplication (MVM) and matrix-matrix multiplication (MMM).
In the case of MVM, the matrix is mapped onto the CIM array, while the vector serves as the array input. 
The multiplication is performed within each cell, with accumulation occurring along the bitlines or at the output side, producing MVM results directly from the array.
\update{Many classic DNN operators, such as fully connected layers and convolutions, can be readily transformed into MMM or MVM operations.
For instance, while convolutional kernels cannot be directly mapped onto the array, the convolution operations can be unrolled into an equivalent matrix-matrix multiplication (MMM).} This equivalent MMM is subsequently mapped and executed on the CIM array, following the standard MMM procedure.


\subsection{\update{CIM Compilation Works for DNN}}
With the increasing attention on CIM, there has been a significant surge in efforts to develop a compilation optimization stack aimed at facilitating the deployment of DNN algorithms across various CIM architectures \cite{siemieniuk2021occ,sun2023pimcomp,qu2024cim,ankit2019puma}. 

Existing compilation optimization approaches for CIM predominantly emphasize scheduling optimizations, such as task mapping, resource allocation, and dataflow scheduling, to fully exploit the static on-chip resources of CIM chips, thereby reducing latency.
For example, OCC \cite{siemieniuk2021occ}, built upon MLIR, utilizes a specific ISA to support scheduling optimization for multiple operators. 
CIM-MLC \cite{qu2024cim} addresses the challenges posed by multi-level and heterogeneous program interfaces in CIM accelerators, implementing weight duplication and pipelining techniques tailored for various CIM computation modes. These compilation strategies alleviate the programming complexity of utilizing CIM processors, accelerate application deployment, and allow researchers to focus more on architecture design.

\begin{figure}[t]
    \centering
    \includegraphics[width=\linewidth]{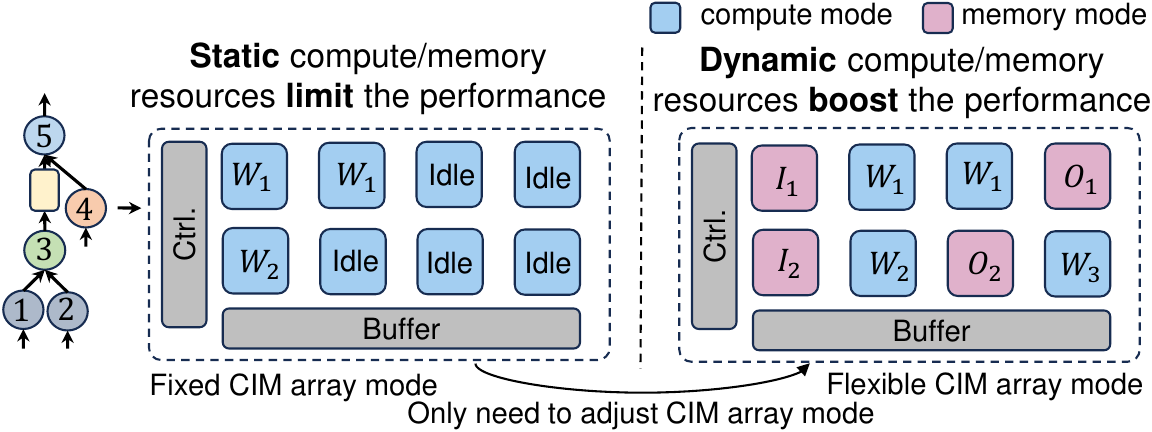}
    \caption{(a) Existing typical CIM mapping method that treats all the CIM arrays as compute arrays; (b) Dual-mode-aware mapping method.}
    \label{fig:motivate}
\end{figure}
However, existing compilers overlook a crucial aspect: the dual-mode capability of CIM arrays, resulting in suboptimal performance.
As shown in \fig \ref{fig:motivate}, when taking the dual-mode feature of CIM arrays into account, the compiler can dynamically allocate compute and memory resources (b). Thus, it can enhance the DNN performance when keeping more data on the chip by switching the CIM arrays to memory mode. Instead, the traditional compiler has to move these data to off-chip memory, incurring extra latency.

In summary, the dual-mode capability of CIM arrays introduces a powerful mechanism for dynamically adjusting on-chip resources to meet the diverse computation and memory demands of various DNN workloads. By intelligently switching CIM arrays between compute and memory modes, we can flexibly allocate resources based on the specific requirements of different DNN inference tasks, ultimately optimizing performance.


\section{Motivation}
\update{This section describes the motivation behind developing a dual-mode-aware DNN compiler for CIM accelerators.
We identify the diverse on-chip computing and memory requirements inherent in real-world DNNs. Additionally, we discuss the opportunities of meeting these application requirements through the optimization of CIM array mode configuration during compilation.}

\subsection{Insights into Diverse DNN Requirements}


\begin{figure}[t!]
    \centering
    \includegraphics[width=\linewidth]{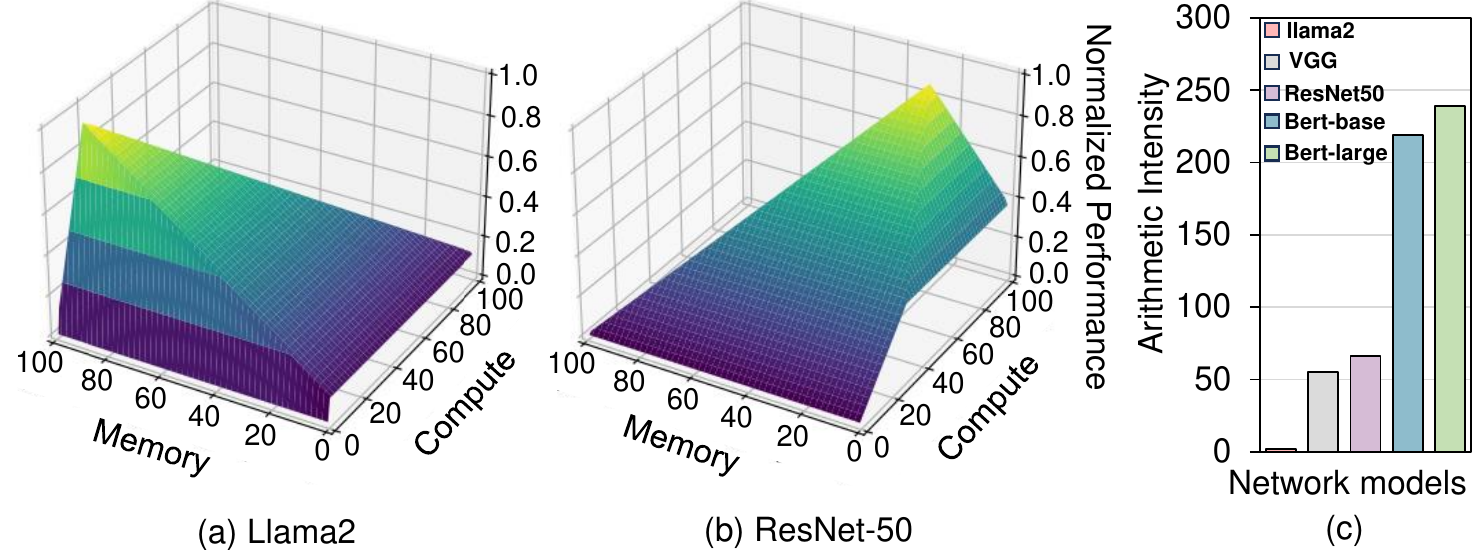}
     \caption{(a)(b) Normalized performance variation with the changes of compute/memory array; (c) Arithmetic Intensity.}
    \label{fig:background2}
\end{figure}

\noindent
\textbf{Variations among different network architectures.}
Mainstream neural networks exhibit diverse architecture designs, leading to varied hardware requirements \cite{he2016deep,simonyan2014very,dosovitskiy2020image,devlin2018bert,redmon2016you,touvron2023llama,nerf,rombach2022high}. 
\fig \ref{fig:background2} (a)(b) illustrates the normalized performance variation heatmap of Llama2 \cite{touvron2023llama} and ResNet-50 \cite{he2016deep} with changes in the number of arrays in compute/memory mode. We assume there is a total of 100 dual-mode CIM arrays on-chip, where the switchable CIM array is as Dynapalsis \cite{kim202316}. 
The memory and compute axes represent the number of CIM arrays in memory and compute mode, respectively.
The vertical axis indicates the theoretical performance normalized to the optimal performance with the same amount of total arrays.
Green indicates better performance, while dark blue indicates poor performance.
Llama2 and ResNet-50 exhibit distinct preferences for hardware resource allocation, stemming from their differing arithmetic intensities. 
 As illustrated in \fig \ref{fig:background2}(c), ResNet-50 has a significantly higher average arithmetic intensity compared to Llama2. 
Consequently, Llama2, with its lower arithmetic intensity, does not require extensive computing resources but necessitates increased on-chip memory to complete its operations. 
Thus, Llama2 demands more CIM arrays in memory mode to meet its computational needs. Conversely, ResNet-50, with its higher arithmetic intensity, benefits from greater compute resources to achieve optimal performance.
Furthermore, \fig \ref{fig:background2}(c) indicates the varying arithmetic intensities across different models. 
Therefore, the dynamic adjustment of memory and compute resources in CIM is essential to provide optimal performance for diverse DNNs.

\begin{figure}[t]
    \centering
    \includegraphics[width=0.9\linewidth]{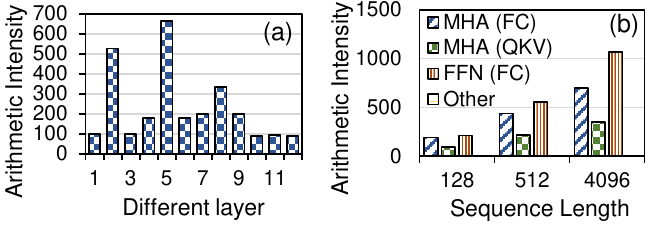}
    \caption{(a) Layer-specific arithmetic intensity of ResNet-50 \cite{he2016deep}; (b)Arithmetic intensity of BERT-Large with different sequence lengths \cite{devlin2018bert}.}
    \label{fig:layer}
\end{figure}

\begin{figure*}[t]
    \centering
    \includegraphics[width=0.95\linewidth]{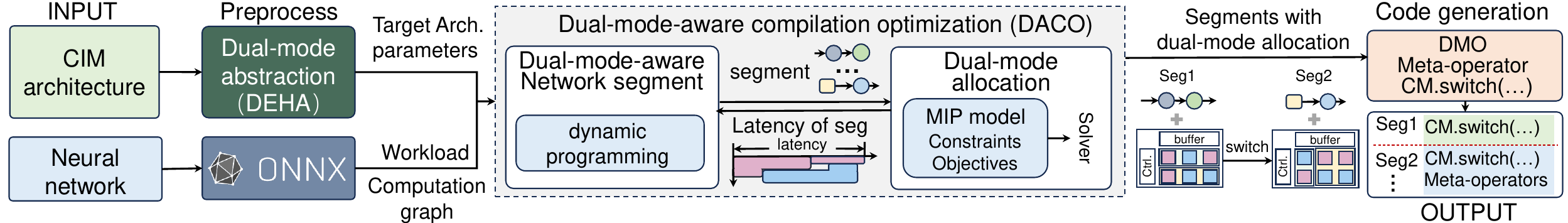}
    \caption{Overview of dual-mode-aware compilation process.}
    \label{fig: workflow}
\end{figure*}


\noindent
\textbf{Layer-wise variations within the same network.} Within the same neural network, different layers also have varying hardware demands due to factors such as input data size and network parameters, including weight kernel size. 
For example, as illustrated in the \fig \ref{fig:layer}(a), ResNet-50\cite{he2016deep} comprises four distinct blocks, each containing three configurations of convolution layers. The arithmetic intensity of these three layers varies significantly, ranging from below $100~\mathrm{FLOPs/MOP}$ to over $700~\mathrm{FLOPs/MOP}$.


\noindent
\textbf{Variations on different workload scales.}
Transformer-based \cite{vaswani2017attention} NLP models, such as BERT \cite{devlin2018bert}, show dynamic resource requirements based on varying input and output sequence lengths.
As illustrated in the \fig \ref{fig:layer}(b), the arithmetic intensity of the model fluctuates significantly with the input and output sequence length, varying from under 150 FLOPs/MOP to over 1000 FLOPs/MOP.
Additionally, different computation stages within the models, such as fully connected (FC) layers and query-key-value (QKV) computations, display varying arithmetic intensities. For example, FC layers demonstrate much higher arithmetic intensity compared to QKV computations as the sequence length increases.
As sequence length grows, more memory is needed to store intermediate states and longer contextual information, while additional computational resources are required to manage the increased complexity of the attention mechanism. Consequently, the demand for compute and memory resources dynamically adjusts with sequence length.

\subsection{Opportunity of CIM Dual-Mode Switch}
\update{Given the diverse resource demands of various DNN architectures, layers, and workload scales, dynamic hardware resource allocation is crucial for optimizing model execution performance. A static compute/memory resource ratio is often insufficient to achieve optimal efficiency across different scenarios. Traditional compilation techniques typically struggle with inefficiencies due to their inability to adapt to fluctuating resource requirements.}
\update{By leveraging the dual-mode switching capability of CIM arrays, we can dynamically alternate between compute and memory modes, enabling CIM accelerators to more effectively accommodate the diverse needs of DNN workloads. Specifically, repurposing compute arrays into memory arrays allows CIM accelerators to expand on-chip memory resources, which is particularly beneficial for storing dynamically generated activations in DNNs.
This flexibility can significantly boost overall system performance and energy efficiency by tailoring hardware configurations to the unique requirements of each DNN model.}

\noindent

To leverage this flexibility, we propose CMSwitch, a dual-mode-aware DNN compiler for CIM processors, ensuring that the dual-mode CIM arrays provide optimal performance for any given workload.
In the following sections, we will introduce the workflow of CMSwitch, elaborating on the dual-mode-aware compilation optimization pass.

\section{Dual-Mode-Oriented Compilation Stack}

\subsection{Overall Workflow}
\fig \ref{fig: workflow} illustrates the workflow of CMSwitch, \update{which takes user-defined hardware parameters and neural network applications as inputs.}
The neural network is initially converted into ONNX format \cite{bai2019}, lowering it to a computation graph expression.
 To integrate compute-memory mode switching into the compilation optimization space, we incorporate the dual-mode functionality of arrays in the hardware abstraction. This is achieved by introducing the methods and overheads associated with compute-memory mode switching into the hardware abstraction parameters.
 
 During compilation optimization, 
to minimize application latency within the joint optimization space, \name develops a divide-and-conquer two-step policy.
\name first decides network segmentation that accounts for dual-mode switch overheads, and then optimizes the dual-mode resource allocation and scheduling for operators within each segment.
Through iterative exploration and optimization using Dynamic Programming (DP) and Mixed-Integer Programming (MIP), \name derives the globally optimal network segmentation schedule, along with the corresponding resource allocation and mapping results for each operator.

Furthermore, to effectively present our compilation results, we introduce meta-operators specifically designed for dual-mode switching. These meta-operators facilitate the output of compilation results that incorporate the compute-memory switch scheme.
Upon obtaining the memory-compute mode switch plan offline, the actual dual-mode switch needs to be executed online with the support of the dual-mode CIM.

\begin{figure}[t]
    \centering
    \includegraphics[width=\linewidth]{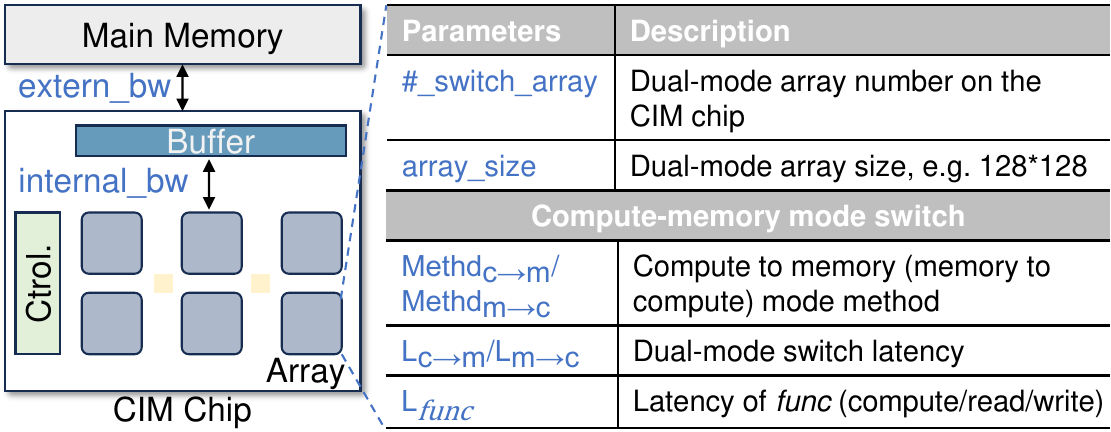}
    \caption{Dual-mode Enhanced Hardware Abstraction.}
    \label{fig: abstraction}
\end{figure}
\subsection{Dual-Mode Enhanced Hardware Abstraction (DEHA)}
Hardware abstraction is crucial to providing essential hardware information to the compiler for the compilation process.
As shown in the \fig \ref{fig: abstraction}, we incorporate the dual-mode parameter of the array and the related switch function into the hardware abstraction. Together with architecture parameters, this abstraction enables \name to access the optimization space of the dual-mode CIM array and relevant architecture parameters.


\noindent
\textbf{Dual-mode CIM architecture.} 
In abstracting the CIM architecture, we model the CIM chip hierarchically. Given our focus on optimizing the dual-mode CIM, we simplify the abstraction to include only two essential tiers: chip and array.
At the finest granularity, our abstraction is at the CIM array level, which represents the smallest hardware unit capable of mode switching.  
Users are required to define parameters such as the number of dual-mode arrays, the array sizes, internal bandwidth, and external global bandwidth, all of which significantly influence the behavior and performance of the CIM chip.

\noindent   
\textbf{Dual mode switch.} 
During the compilation optimization process, it is crucial to consider both the functionality and overhead of the dual-mode switch to enable the compiler to balance the benefits of mode switching and explore optimal performance results.
To facilitate the compilation process, it is essential to define the method and overhead of the compute-memory mode switch adopted by the target chip at the hardware abstraction stage. This may involve techniques such as altering the inputs of wordlines or bitlines. Additionally, the overhead of the compute-memory mode switch is assessed at the granularity of the switchable arrays. These parameters significantly impact subsequent compilation optimization solutions.

When CIM arrays operate in different modes, executing a single operation—such as activating an array for computation or data reading—may incur varying overheads. To account for this, we introduce an option in the hardware abstraction parameters to record the overhead of computation and memory operations. 
This approach ensures that the associated overheads of read-write and computation of arrays are considered during the compilation optimization process, thereby evaluating the impact of different modes on overall performance.
\subsection{Dual-Mode-Aware Compilation Optimization}
Based on the abstraction of the hardware architecture, we proposed the Dual-Mode Aware Compilation Optimization (DACO) to allocate the dual-mode CIM arrays for each CIM-supported operator, to minimize overall execution latency.
This phase gets the ONNX-format DNN and hardware abstraction as input. 
We employ a divide-and-conquer two-step policy.
First, dynamic programming (DP) is utilized to partition the network into multiple segments, taking into account the switching overhead between segments.
Following segmentation, mixed linear programming (MIP) is applied to co-optimize the allocation of on-chip computing/memory resources and scheduling for operators, adhering to the constraint of total available resources. 
During this optimization process, the dual-mode CIM arrays are dynamically adjusted to either memory or compute mode based on the optimization objective. 
Finally, this phase outputs the network segmentation results along with the corresponding allocation of dual-mode CIM array in memory and compute mode for each segment. 
The notation we used for optimization space formalization in this section is summarized in \tab \ref{tab:not}. 

\subsubsection{Dual-Mode-Aware Network Segment}
\begin{table}[]
\caption{Notations Used in Dual-mode Compilation.}
    \label{tab:not}
\resizebox{\linewidth}{!}
{

\begin{tabular}{l|l}
\hline
          \rowcolor{mygray}\multicolumn{2}{c}{Variables - decide during optimization} \\ \hline
         $\lambda_{min}(i,x,y)$ &  \parbox[t]{9cm}{taking value 1 if CIM array (x,y) is assigned to $O_i$ in memory mode as input buffer, 0 otherwise.}       \\ \hline
         $\lambda_{mout}(i,x,y)$ &  \parbox[t]{9cm}{taking value 1 if CIM array (x,y) is assigned to $O_i$ in memory mode as output buffer, 0 otherwise.}       \\ \hline
         $\lambda_{c}(i,x,y)$ & \parbox[t]{9cm}{taking value 1 if CIM array (x,y) is assigned to $O_i$ in compute mode, 0 otherwise.}        \\ \hline
         $Mem_{O_i}$ & \parbox[t]{9cm}{the number of CIM array in memory mode that $O_i$ has. $Mem_{O_i} = \textstyle\sum_{x,y} \lambda_{min}(i,x,y)+\sum_{x,y} \lambda_{mout}(i,x,y)$ } \\ \hline
         $Com_{O_i}$ & \parbox[t]{9cm}{the number of CIM array in compute mode that $O_i$ has. $Com_{O_i} =\textstyle\sum_{x,y} \lambda_{c}(i,x,y)$}        \\ \hline
         $Switch_{m\to c}$ & \parbox[t]{9cm}{arrays from memory mode to compute mode for segment S' to S, $Switch_{m\to c} =\textstyle \sum_{x,y}(\sum_{O_i \in S'}\lambda_{m}(i,x,y) \odot \sum_{O_j \in S}\lambda_{c}(j,x,y))$, where $\lambda_{m}(i,x,y)=\lambda_{min}(i,x,y)+\lambda_{mout}(i,x,y)$}        \\ \hline
         $Switch_{c\to m}$ & \parbox[t]{9cm}{arrays from compute mode to memory mode for segment S' to S, $Switch_{m\to c} =\textstyle \sum_{x,y}(\textstyle\sum_{O_i \in S'}\lambda_{c}(i,x,y) \odot \textstyle\sum_{O_i \in S}\lambda_{m}(j,x,y))$}        \\ \hline
          \rowcolor{mygray}\multicolumn{2}{c}{Constants - determine by application and CIM chip initially} \\ \hline
         $IN_{O_i}$/$OUT_{O_i}$ & input data/output data of operator $O_i$        \\ \hline
         $AI_{O_i}$ & arithmetic intensity  of operator $O_i$        \\ \hline
         $N_{cim}$ & the number of dual-mode switchable CIM array         \\ \hline
         $OP_{cim}$ & operation/cycle a CIM array can provide, $OP_{cim} \propto array\_size$         \\ \hline
         $D_{cim}$ &  \parbox[t]{9cm}{data/cycle a CIM array can provide in the memory mode, \update{which is impacted by architecture design and user-defined topology.}}\\ \hline
         $D_{main}$ & \parbox[t]{9cm}{data/cycle main memory and original on-chip buffer can provide, $D_{main} \propto extern\_bw + internal\_bw$}\\ \hline
\end{tabular}}
\end{table}
\begin{figure}[t]
    \centering
    \includegraphics[width=\linewidth]{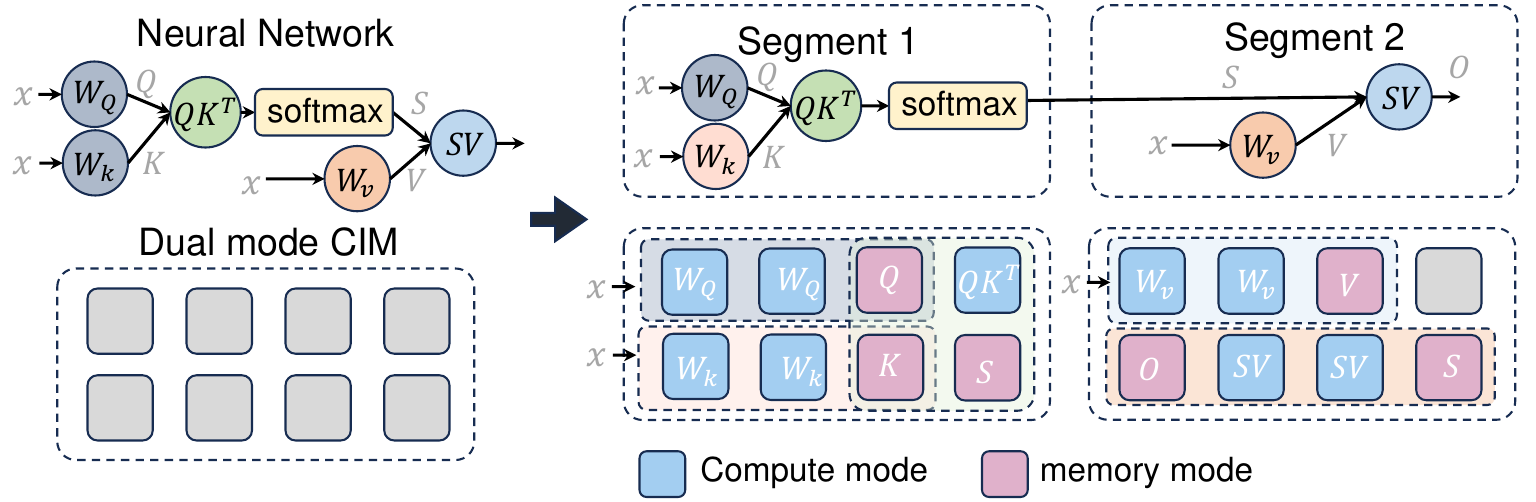}
    \caption{Illustration of Network Segment.}
    \label{fig:seg1}
\end{figure}
In the network segment step, we organize the network compute operators at the granularity of network segments, being allocated on the dual-mode CIM arrays holistically.
Network segmentation offers two key benefits: 1) it reduces the optimization space from an exponential size, considering the entire number of operators, to a more feasible one, and 2)  it facilitates the accommodation of real-world DNN networks that cannot fit entirely on the chip to be optimized on the dual-mode CIM.
As shown in the \fig \ref{fig:seg1}, taking the attention layer as an example, if we segment the attention process into two segmentations, we will execute them serially.
The first segment maps to the hardware and executes, followed by the second segment after the first completes.
Both of them will have their resource allocation plan.
To further refine the segmentation search space, we employ a dynamic programming approach to optimize network latency in the context of scheduling network segments.

For a network $N$ with $m$ CIM-supportable operators (e.g., MVM and MMM), we first topologically sort these operators, denoted as $O_1, O_2, ..., O_m$, where $O_i$ and $O_j$ ($i,j \in \{1,2,...,m\}$) satisfy that if $i < j$, then $O_i$ is completed no later than $O_j$. The dependency relationship between operators is denoted as $W$, where $w_{i,j } \in W$ indicates that the output of $O_i$ is input to $O_j$. 
The segments after segmentation are denoted as $S_{i,j}$, indicating that operators from $O_i$ to $O_j$ belong to the same segment $S_{i,j}$. 
\update{For operators that cannot fit directly onto the CIM accelerator, we will partition them into smaller sub-operators. This partitioning process uses a greedy strategy, with the partition granularity determined by the available on-chip resources. This approach ensures that each resulting sub-operator can be fully mapped onto the chip. Finally, we replace the original operator in the sorted operator list with these sub-operators, enabling efficient execution on the CIM accelerator.}
To minimize inference latency, the segmented network execution overhead includes both intra-segment execution latency and inter-segment mode-switch latency.

\noindent
\textbf{Intra-segment latency.} Within each segment, all operators are mapped to the CIM chip simultaneously. Therefore, optimization methods such as pipelining can be employed to minimize the overall latency within the segment. The introduction of dual-mode CIM arrays makes the intra-segment latency highly dependent on on-chip memory and computation resource allocation. To optimize the intra-segment overhead, we will detail the dual-mode-aware resource allocation algorithm in the following subsection. We denote the intra-segment latency with resource allocation plan $A$ as $T_{i,j}^{intra}(A)$ for $S_{i,j}$.

\begin{figure}[t]
    \centering
    \includegraphics[width=\linewidth]{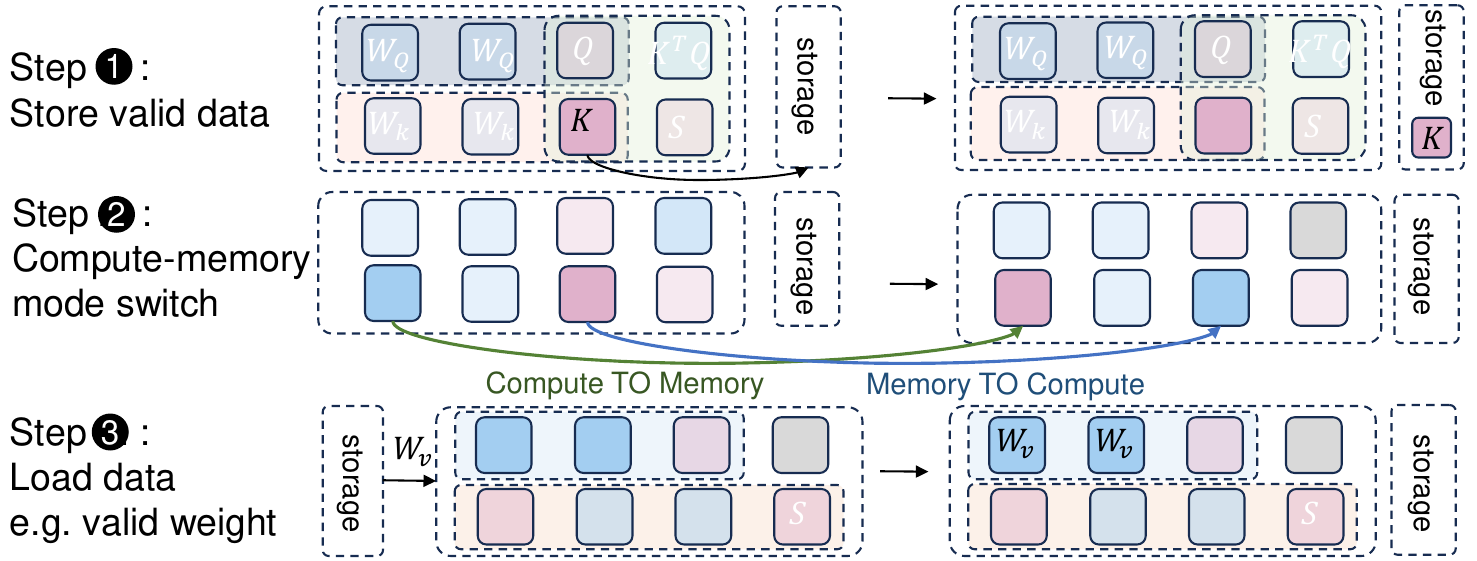}
    \caption{Three sources of inter-segment overhead.}
    \label{fig:seg2}
\end{figure}

\noindent
\textbf{Inter-segment latency.} Inter-segment latency encompasses the latency caused by switching CIM memory and computation modes and on-chip and off-chip data swapping. 
Specifically, as illustrated in \fig \ref{fig:seg2}, the inter-segment mode switch mainly consists of three steps, storing the valid on-chip data to storage, performing the mode switch between the compute/memory, and loading data. 
We formalize the inter-segment latency based on the two adjacent segments, $S_{k,i-1}$ with the resource allocation plan $A'$ and $S_{i,j}$ with the resource allocation plan $A$ ($k< i\leq j$).

For step one: if segment $S_{k,i-1}$ contains more memory arrays than $S_{i,j}$, and the output data from the segment $S_{k,i-1}$ is needed for subsequent computations, this data must be written back to the main memory before switching this CIM arrays from memory mode to compute mode. We denote the data store latency as $T_{i-1,i}^{wb}(A',A)$, and it can be estimated according to the data transfer volume and the memory bandwidth.
For data that can be processed in place and will not be reused, such as softmax results in attention, the corresponding CIM arrays can be directly switched from memory to compute mode without data write-back.

For the second step, when switching between memory and compute modes, if the required arrays differ between adjacent network segments, the dual-mode switch overhead must be considered. When the $S_{k,i-1}$ uses arrays in compute mode while the $S_{i,j}$ requires them in memory mode, the overhead of switching those arrays from compute mode to memory mode must be accounted for, and vice versa. The overhead $T_{i-1,i}^{swc}(A',A)$ is detailed in the Eq. \ref{eq:swc}\
\begin{equation}
    \scriptsize 
    T_{i-1,i}^{swc}(A',A) =
        L_{M\to C}\times Switch_{m \to c} +
        L_{C\to M}\times Switch_{c \to m}.
    \label{eq:swc}
\end{equation}

For the third step, as different segments process different operators, the weights stored in the compute arrays must be reloaded and rewritten. 
Each operator has unique weights, necessitating an update of the compute arrays' stored weights accordingly.
The overhead $T_{i-1,i}^{rw}(A',A)$ is shown in the Eq. \ref{eq:1}:
\begin{equation}
  T_{i-1,i}^{rw}(A',A) =\textstyle \max_{O_l \in S_{i,j}}{Com_{O_l}\times Latency\_write}.
  \label{eq:1}
\end{equation}

Based on the intra- and inter-segment overhead, network segmentation can be formalized as the following dynamic programming problem.
As shown in Eq. \ref{eq:dy}, we denote the best network segment solution with the corresponding resource allocation plan $A^*$ as the $L[m][A^*]$. The total cost of the network from operator 1 to $m$ is the sum of the costs of two segments: from 1 to $i-1$ and from $i$ to $m$, plus the mode-switch latency between these segments. To reduce the solution space, any network segment exceeding the total on-chip resources is deemed invalid. 
If a segment requires more compute and memory arrays than the available on-chip resources, it must be further segmented during execution.
\begin{equation}
    L[m][A^*] = \min_{1 \leq i < m} \{L[i][A'] + T_{i, m}^{intra}(A) + T_{i-1, i}^{inter}(A',A)\}, \\
    \label{eq:dy}
\end{equation}
where $T_{i-1, i}^{inter}(A',A)$ equals to 
\begin{equation}
    T_{i-1, i}^{inter}(A',A) = T_{i-1, i}^{wb}(A',A)+T_{i-1, i}^{swc}(A',A)+T_{i-1, i}^{rw}(A',A).
    \label{eq:inter}
\end{equation}
By traversing all potential choices and backtracking the segmentation plan according to Eq \ref{eq:dy}, we can efficiently find the optimal segmentation strategy, thereby minimizing execution time. This dynamic programming approach significantly reduces the search space complexity compared to exhaustive search methods.

\subsubsection{Unified Dual-Mode Allocation with Scheduling}
In this section, we formalize the resource allocation and operator scheduling co-optimization problem as a Mixed-Integer Programming (MIP) problem to minimize the execution latency.
For each network segment, the allocation optimization aims to determine the optimal modes of CIM arrays assigned to each operator. Operators within the segment are scheduled in the pipelined fashion to maximize computing parallelism. To capture the interdependence of compute and memory allocation and the scheduling for operators within a segment, we define specific objectives and constraints, where constraints dictate dual-mode CIM resources available for operators and objectives that optimize latency considering the pipeline structure, respectively.
The details of our formulation are explained in the following contents.

Here we use the 2-d coordinate $(x,y)$ to indicate the CIM arrays in the chip and use the $\lambda_{z}(i,x,y), z \in \{min, mout, c\}$ to indicate the resources allocation and mapping for operator $O_i$, 
where $min$/$mout$ means array in memory mode as input/output buffer, respectively, and $c$ means array in compute mode. 
When the CIM array (x,y) assigned to $O_i$ is in memory mode as input and output buffer, $\lambda_{min}(i,x,y)=1$ and $\lambda_{mout}(i,x,y)=1$, otherwise they are 0. Similarly, when the CIM array (x,y) assigned to $O_i$ is in compute mode, $\lambda_{c}(i,x,y)=1$.
Our goal is to find the optimal $\lambda_{z}(i,x,y)$ for each segment under the constraints imposed by the resource allocation and the objective related to the pipeline scheduling strategy.

\begin{figure}[t]
    \centering
    \includegraphics[width=\linewidth]{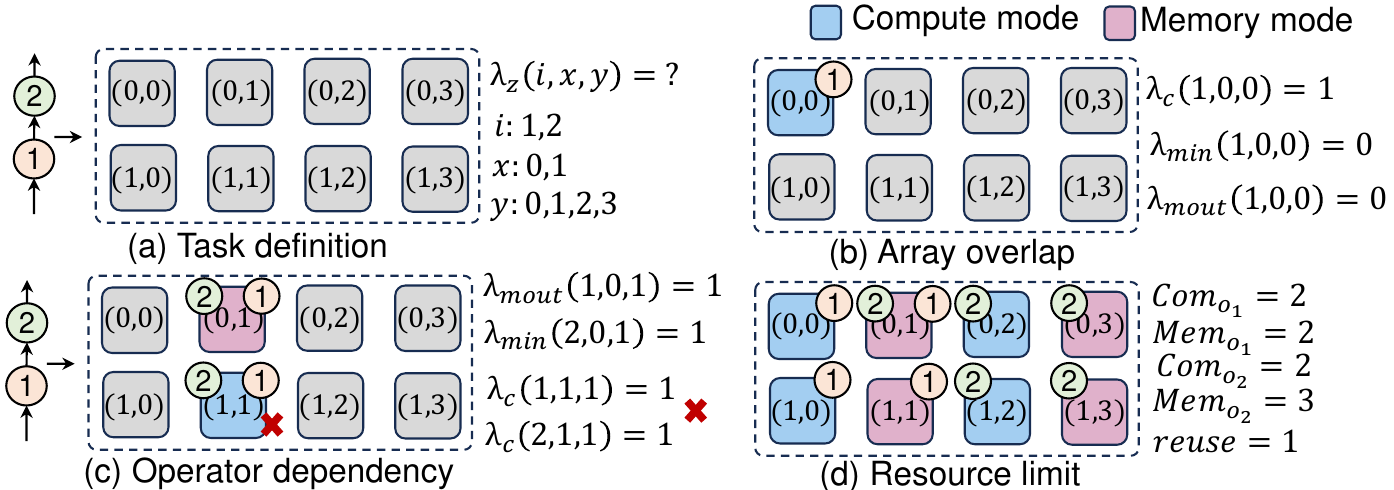}
    \caption{Constraints illustration for resource allocation.}
    \label{fig:cons}
\end{figure}
\begin{figure}[t]
    \centering
    \includegraphics[width=\linewidth]{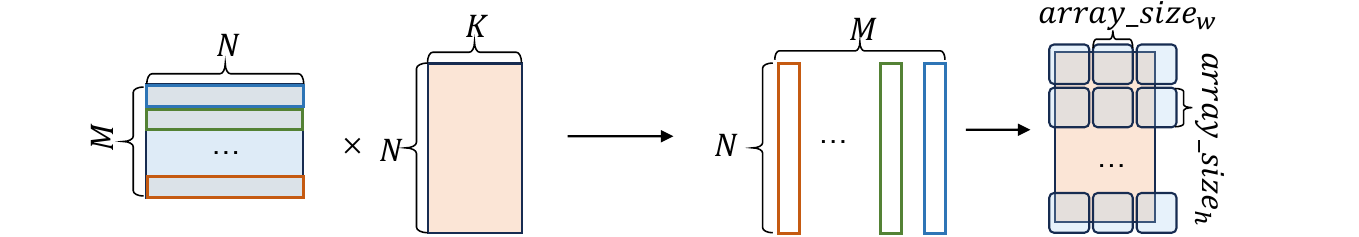}
    \caption{Example of MMM.}
    \label{fig:mip}
\end{figure}
\noindent\textbf{Constraints.}
As the dual-mode resource space is defined, some optimization rules should be considered to include only the valid CIM array allocation solutions.
Here, we use $S_*$ to indicate any possible network segment. 

\textit{1) Array overlap.} A CIM array can be either memory or compute mode for an operator, but not both. Therefore, for each CIM array $(x,y)$, if it is allocated to operator $O_i$, only one of $\lambda_{min}(i,x,y), \lambda_{mout}(i,x,y)$, or $\lambda_{c}(i,x,y)$ can be 1.
\begin{equation}
    \scriptsize\forall O_i \in S_*, \forall x,y: \\
    \lambda_{min}(i,x,y)+\lambda_{mout}(i,x,y)+\lambda_{c}(i,x,y) \leq 1.
\end{equation}
As shown in \fig \ref{fig:cons}(b), CIM array $(0,0)$ is assigned to $O_1$ in compute mode, so it can not be a memory array.

\textit{2) Operator dependency.} If one operator's output $O_i$ serves as the input of the next $O_j$, the output memory arrays of $O_i$ can directly serve as the input memory arrays of $O_j$. Thus, a portion of the output memory buffer for $O_i$ can be reused as an input memory buffer for $O_j$.
\begin{equation}
\scriptsize
\begin{split}
    &\forall O_i,O_j \in S_*, if\space w_{i,j} \in W: \\
    &\textstyle \sum_{x,y} \lambda_{mout}(i,x,y) \cdot \lambda_{min}(j,x,y) < \frac{OUT_{o_i} \bigcap IN_{o_j}}{array\_size}.
\end{split} 
\end{equation}
As the example shown in \fig\ref{fig:cons}(c), 
$O_1$'s output is the input of $O_2$, where array $(0,1)$ assigned to $O_1$ as output buffer can be input buffer of $O_2$.

Otherwise, there are no reused CIM resources between the adjacent operators. 
\begin{equation}
\begin{split}
    &\forall O_i,O_j \in S_*, w_{i,j} \notin W, z\in\{mout,min,c\}: \\
    &\textstyle \sum_{x,y} \lambda_{z}(i,x,y) \cdot \lambda_{z}(j,x,y) = 0 .
\end{split} 
\end{equation}
As shown in \fig \ref{fig:cons}(c), CIM array $(1,1)$ can not be compute array for operator $O_1$ and $O_2$ at the same time.

\textit{3) Resource limit.} The total number of CIM arrays assigned to all operators must not exceed the available resources. $H_{i,j}(x,y)$ denotes $\lambda_{mout}(i,x,y) \cdot \lambda_{min}(j,x,y)$. 
\begin{equation}
    \textstyle \sum_{O_i \in S_*}(Mem_{O_i}+Com_{O_i}) - \textstyle \sum_{i,j} \textstyle \sum_{x,y} H_{i,j}(x,y) < N_{cim}.
\end{equation}
As shown in \fig \ref{fig:cons}(d), the allocated arrays are under the resource limit.

\noindent\textbf{Objective function.}
Our objective is to optimize the allocation of arrays in memory or compute mode to minimize the execution latency of the network segment under a pipeline scheduling strategy. 
Through the pipeline, operators can execute in parallel.
Thus, the latency of the segment can be approximated as the maximum execution time of any single operator within that segment. 
We denote the latency of $O_i$ as $L_{O_i}$ and get the following objective function:
\begin{equation}
    \label{eq:objct}
    \min \max L_{O_i}, \forall O_i \in S_* .
\end{equation}

The system performance cost model for $L_{O_i}$ estimates off-chip data access latency and on-chip computation latency based on the allocated compute and memory arrays.
To quickly estimate the latency $L_{O_i}$, we develop a latency model as shown in Eq.~\ref{eq:latency}. 
It is a function of the number of allocated compute and memory arrays ($Com_{O_i}$ and $Mem_{O_i}$).
When we allocate $Com_{O_i}$ compute arrays for $O_i$, they support $\scriptsize C=Com_{O_i} \cdot OP_{cim}$ computation amount per cycle, where $OP_{cim}$ denotes the computation amount per cycle a CIM array can provide. 
When we allocate $Mem_{O_i}$ memory arrays, they can access $Mem_{O_i} \cdot D_{cim}$ data per cycle, where $D_{cim}$ represents the data per cycle a CIM array can provide. Combined with the data from storage and the original buffer ($D_{main}$), the accessible data per cycle is $Mem_{O_i} \cdot D_{cim} + D_{main}$. Given the arithmetic intensity of $O_i$ ($AI_{O_i}$), this supports $\scriptsize M=(Mem_{O_i} \cdot D_{cim} + D_{main}) \cdot AI_{O_i}$ computation amount.
The smaller value between the $C$ and $M$ determines the effective computation amount per cycle.
$L_{O_i}$ can be estimated by the total computation amount ($OP_{O_i}$) and the $min(C,M)$.
Therefore, when $AI_{O_i}$ is large, it is preferable to allocate more compute arrays. Conversely, when $AI_{O_i}$ is small, it is better to allocate more memory arrays.

\begin{equation}
    \scriptsize L_{O_i} \propto \frac{OP_{O_i}}{\min (Com_{O_i} \cdot OP_{cim}, (Mem_{O_i}\cdot D_{cim} + D_{main})\cdot AI_{O_i} )}. 
\label{eq:latency}
\end{equation}
Taking matrix-matrix multiplication(MMM) as an example, as shown in \fig \ref{fig:mip}, for an operator $O_i$, the computation amount a CIM array can provide is: $\scriptsize \frac{N\times K}{\lceil\frac{N}{array\_size_h}\rceil\times\lceil\frac{K}{array\_size_w}\rceil}$.
Meanwhile, $N$ data can support $N \times K$ MAC computations, which means $AI_{O_i} = K$. 
The total number of multiply - accumulate operations needed is $OP_{O_i} = M \times N \times K$. 


Using the optimization solver Gurobi \cite{gurobi}, we solve the most efficient memory/compute mode allocation for each network segment, thereby optimizing the overall performance of the dual-mode CIM processor. 

\begin{algorithm}[t!]
  \caption{Summary of DACO.}
  \label{alg:sum}
\begin{algorithmic}[1]
    \STATE {\bfseries Input:} Neural network $N$ in ONNX format and CIM hardware abstraction.
    \STATE {\bfseries Output:} The network segment $S$ and the corresponding dual-mode CIM array allocation $A^*$ for each segment.
    \STATE \emph{\textcolor{gray}{// Preprocess graph.}}
    \STATE $(O_1, \ldots, O_m) \leftarrow \text{Flatten}(G)$
    \STATE \emph{\textcolor{gray}{// Run the network segment dynamic programming}}
     \STATE L[0][$\cdot$] $\leftarrow$ 0
    \FOR{$1 \le i \le j \le m$}        
        \STATE \emph{\textcolor{gray}{//Impossible cases are skipped to reduce search space}}
        \IF{min \# of CIM array $S_{i,j}$ required $< N_{cim}$}
            \STATE  \emph{\textcolor{gray}{// Run the resources allocation.}}
            \STATE $T_{i,j}^{intra}(A)\leftarrow MIP(S_{i,j})$ according to Eq \ref{eq:objct}. 
        \ELSE
            \STATE $T_{i,j}^{intra}(\cdot)\leftarrow \infty$
        \ENDIF
        \STATE Compute $T_{i-1,i}^{inter}(A',A)$ according to Eq \ref{eq:inter}.
        \STATE L[j][A] $\leftarrow$ min(L[i][A$'$] + $T_{i,j}^{intar}(A)$ + $T_{i-1,i}^{inter}(A',A)$, L[j][A])  
        \STATE S$_{record}$.update(i,j)
    \ENDFOR
    \STATE L$_{min}$ = $\min$ L[m][$A^*$]
    \RETURN backtrack(L, S$_{record}$, L$_{min}$) for network segment plan S and corresponding $A^*$.
\end{algorithmic}
\end{algorithm}
Algorithm~\ref{alg:sum} summarizes the dual-mode-aware compilation optimization (DAMO) workflow, detailing the interaction between network segment and resource allocation. \update{Once the network segment and CIM array allocation plans are established, we perform post-allocation optimization, such as weight duplication, commonly used in CIM compilation optimization \cite{qu2024cim}, to further enhance kernel mapping.}

\subsection{Dual-mode Support in Meta-Operator (DMO)}
\begin{figure}[t]
    \centering
    \includegraphics[width=\linewidth]{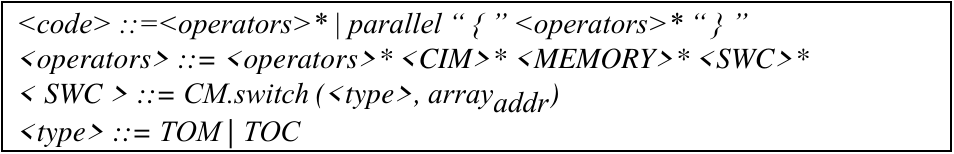}
    \caption{Syntax of dual-mode enhanced code generation.}
    \label{fig:syntax}
\end{figure}

Once the frontend optimization outputs the network segmentation and the allocation strategy of memory/compute resources, 
\name uses meta-operator flow to express the compilation result.
To accommodate the diverse methods for performing these switches, we use a meta-operator flow rather than machine code for better generality. Additionally, it can be integrated into other backends \cite{qu2024cim}. 
The introduced meta-operators and their corresponding syntax are shown in \fig \ref{fig:syntax}. 
We add the $CM.switch$ operator, which supports two types, $TOM$ and $TOC$, representing mode switching at the granularity of CIM arrays. 
When the meta-operator type is $TOM$/$TOC$, it means switching the $array_{addr}$ to memory/compute mode.
When a CIM array is converted to memory mode, it is marked as a valid memory unit, effectively serving as an on-chip buffer. Alongside the new dual-mode switch meta-operators, we use standard operators to represent normal computation and memory access. 
Additionally, we use $parallel \{\}$ to represent the network segment as operators in a segment will execute in parallel.
\update{Users can convert the meta-operator flow into ISA or machine code suitable for their specific CIM chips.}
\section{Evaluation}
\subsection{Setup}
\begin{figure*}[t]
    \centering
    \includegraphics[width=\linewidth]{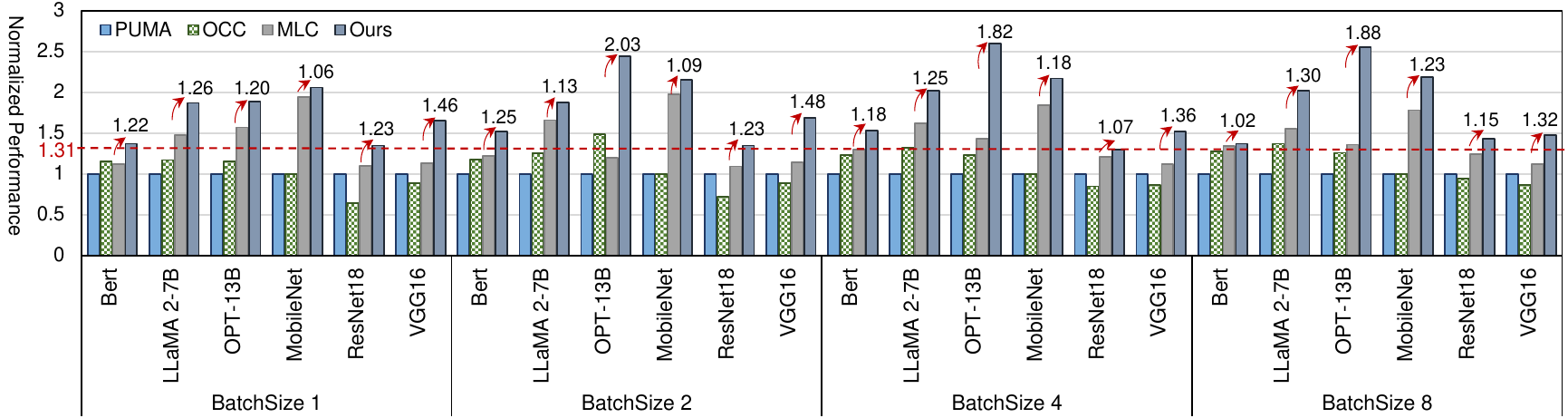}
    \caption{\update{Speedup compared to the baselines. The red arrows with numbers highlight the speedup of CMSwith compared to the main baseline, CIM-MLC\cite{qu2024cim}. The red line highlights the Geomean bar of performance improvement over CIM-MLC.} }
    \label{fig:data}
\end{figure*}

\noindent\textbf{Simulator.}
For the functional simulation, we adopt the CIM-MLC functional simulator \cite{qu2024cim}. We use the functional simulator to execute the generated meta-operator flows within the CIM architecture.
By comparing the execution result with the PyTorch framework \cite{paszke2019pytorch}, we verify the effectiveness of our compilation results.
To evaluate execution latency, we built our simulator upon existing open-source simulators \cite{chen2018neurosim,xia2017mnsim}, incorporating necessary modifications to simulate the dual-mode switch. We also modified the hardware configuration to align with the specifications of Dynaplasia \cite{kim202316}, our target hardware.

\noindent\textbf{CIM architecture configuration.}
Our target CIM architecture configuration is based on Dynaplasia \cite{kim202316}, a real CIM chip that supports the compute-memory mode switch. 
The main parameters of the architecture are listed in the \tab \ref{tab:config}. The CIM mode switch of Dynaplasia is achieved by altering the global wordline input.
Consequently, the actual execution of the $CM.switch$ operator at the runtime is setting the corresponding signal to the global wordline.

\noindent\textbf{Network benchmark.}
To verify both the efficiency and generality of CMSwitch, we use various types of neural networks as our network benchmark. For convolution-based architecture, we use the classic MobileNet \cite{sandler2018mobilenetv2}, ResNet \cite{he2016deep}, and VGG \cite{simonyan2014very} series, tested on the ImageNet dataset\cite{deng2009imagenet}. For the Transformer-based architecture, we use the encode-only model BERT \cite{devlin2018bert} and decode-only models OPT \cite{zhang2022opt} and LLaMA 2 \cite{touvron2023llama}. All models are quantized with 8-bit precision for weights and activations.

\noindent\textbf{Baseline.}
To evaluate the benefits of dual-mode switching for application execution, we compare the compilation results of 
\name with existing CIM compilation works. We adopt three compilers with different optimization strategies as baselines to demonstrate the effectiveness: PUMA \cite{ankit2019puma} (focusing on operator duplication and pipeline scheduling), OCC \cite{siemieniuk2021occ} (optimizing operator mapping via tiling and loop unrolling), and CIM-MLC \cite{qu2024cim} (employing multi-grained pipelining and operator duplication for diverse architecture). Our primary comparison metric is the execution latency of the compiled applications.

\begin{table}[]
    \centering
    \caption{CIM Architecture Configuration.}
    \label{tab:config}
    \resizebox{0.85\linewidth}{!}
{
    \begin{tabular}{c|c}
    \hline
    Parameter & Configuration\\\hline
   $ \#\_switch\_array$     &  96\\
   $ array\_size$     & 320 $\times$ 320\\
   $ buffer\_size$ & $ 10KB\times$8\\
    $internal\_bw$ & 32b/cycle\\
    Method$_{c \to m}$ / Method$_{c \to m}$ & change the input of global IA and IA'\\
    L$_{c \to m}$ / L$_{c \to m}$ & 1 cycle \\\hline
     \end{tabular}}    
\end{table}


\subsection{End-to-End Performance}

\update{We first evaluated the end-to-end performance of CMSwitch, with the results presented in Figure \ref{fig:data}. For transformer-based models, we set the sequence length to 64. Compared to the main baseline, state-of-the-art compilation work CIM-MLC \cite{qu2024cim}, \name demonstrated performance improvements ranging from 1.02$\times$ to 1.25$\times$ (average 1.17$\times$) for the encode-only BERT-large model. For the decode-only LLaMA2-7B model, \name yielded a performance gain of 1.13$\times$ to 1.30$\times$ (average 1.24$\times$). For OPT-13B, \name outperformed the baseline by 1.20$\times$ to 2.03$\times$ (average 1.73$\times$). For CNN models, \name delivered a performance boost of 1.06$\times$ to 1.23$\times$ for MobileNet, 1.07× to 1.23$\times$ for ResNet18, and 1.32$\times$ to 1.48$\times$ for VGG-16. Overall, \name achieved an average speedup of 1.31$\times$, with a maximum improvement of 2.03$\times$ compared to CIM-MLC.}

\update{The fundamental enhancement brought by CMSwitch lies in its ability to expand the compilation optimization space for CIM accelerators. By leveraging the dual-mode switchable capabilities of CIM, CMSwitch dynamically adjusts the balance between computation and memory resources based on the application's requirements, offering more flexible optimization options than all the baselines. This adaptability is particularly advantageous for DNN applications with substantial memory demands and varying arithmetic intensity.}

\update{Specifically, for large transformer-based benchmarks, where the hardware may be unable to fully map the weights, \name optimizes network segmentation by accounting for the overhead associated with switching between compute and memory modes. Meanwhile, combined with a tailored allocation of dual-mode CIM resources, \name allows some arrays to operate in memory mode.}
\update{In contrast, prior compilation frameworks have not adequately considered this issue, leading to under utilization of hardware resources.}
\update{Although acceleration benefits for certain CNNs may be less pronounced, the introduction of the resource-switching capabilities enables better handling of the diverse computational and memory demands across different CNN layers.}
CMSwitch’s dual-mode-aware compilation optimization ensures more efficient handling of resource demands.

The results demonstrate CMSwitch’s versatility across a variety of neural networks with differing hardware resource requirements. It effectively adjusts CIM array modes based on workload characteristics, consistently showing advantages across various batch sizes. This adaptability reflects CMSwitch's capability to optimally allocate compute and memory resources, ensuring performance acceleration at different scales and workloads.


\begin{figure}[t]
    \centering
    \includegraphics[width=\linewidth]{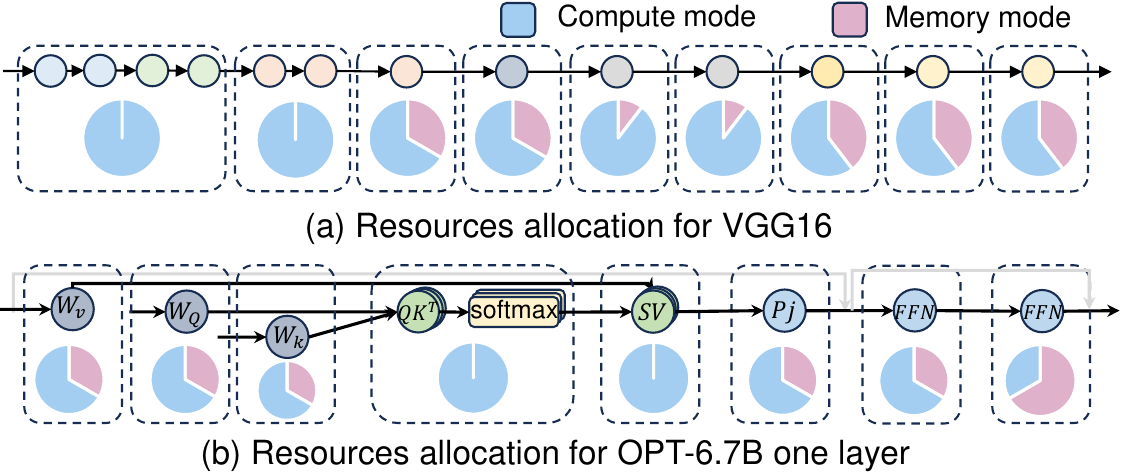}
    \caption{Resources allocation for applications.}
    \label{fig:allocation}
\end{figure}
\begin{figure*}[t]
    \centering
    \includegraphics[width=\linewidth]{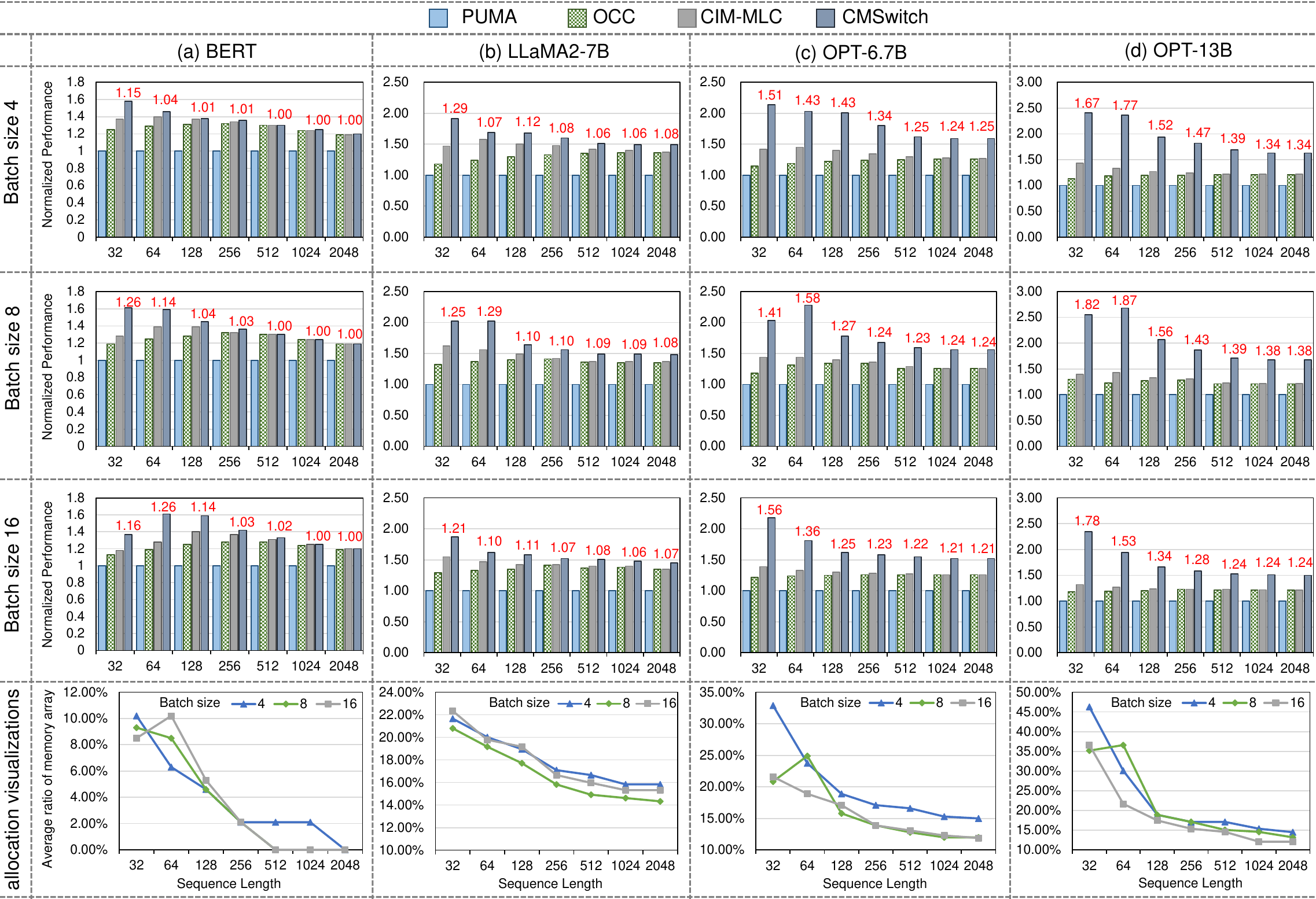}
   \caption{\update{Effectiveness for various workload scales. The horizontal header indicates the model and the vertical header indicates the batch size. (top three rows) Visualization of the memory/compute allocation sensitivity across different sequence lengths for the transformer-based model. (last row) The number in red indicates the speedup compared to the CIM-MLC. }}
    \label{fig:scale}
\end{figure*}

\begin{figure}[t]
    \centering
    \includegraphics[width=\linewidth]{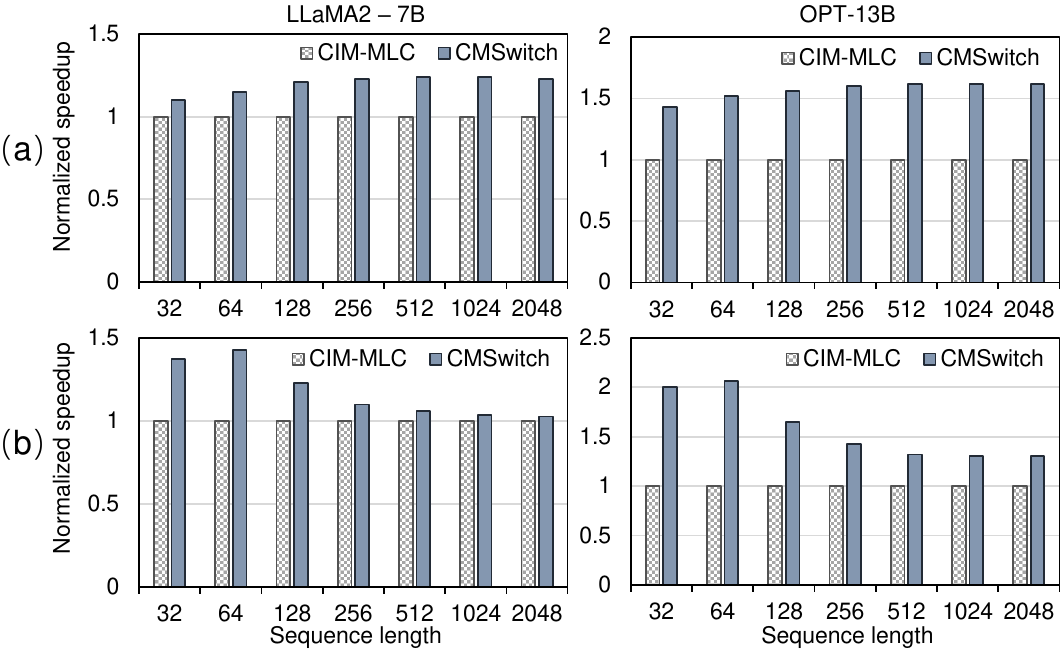}
    \caption{\update{Effectiveness for generative models. (a) Fixed input sequence length; (b) Fixed output sequence length.}}
    \label{fig:scale_vis}
\end{figure}

\subsection{Dual-mode Switch Result Demonstration}
This section presents the allocation of compute/memory arrays after compilation.
As shown in \fig \ref{fig:allocation}, the dashed boxes show the network segment, and the pie charts show the proportion of compute/memory arrays allocated.

For the convolution-based VGG16 (\fig \ref{fig:allocation}(a)), the segmentation results divided the topologically sorted convolution operators 1-4 and 5-6 into two segments, while each of the remaining convolution operators was placed in its own segment. This segmentation aligns with intuitive expectations.
Due to the increasing number of feature map channels in VGG, 
the earlier layers require fewer compute arrays for weight mapping compared to the later layers, making it feasible to group multiple operators into one segment for pipeline parallelism.
Meanwhile, our MIP-based compute/memory allocation strategy provides customized hardware resource support for different network segments. 
As shown in the \fig \ref{fig:allocation}(a), the compiled results allocate more compute arrays for the earlier convolutional operators,
facilitating parallel computation of multiple operators.
With fewer channels, the amount of data required to be loaded is relatively small, and the original on-chip buffer can support the data transfer needs, thus more arrays are used in compute mode.
For the later network layers, especially the final convolutional layers, with more input channels, 
more data is needed in a single MAC computation. Therefore, our model tends to set some CIM arrays to memory mode, providing greater bandwidth for data retrieval.

For the transformer-based model OPT-6.7B, the allocation of compute/memory arrays for one layer is shown in the \fig \ref{fig:allocation} (b). In the standard matrix multiplication, like QKV generation and feed-forward net (FFN), \name allocates 33\%\textasciitilde 67\% CIM arrays in the memory mode based on the operators' demands. 
For example, the last FFN operator needs more input data for each computation and is therefore allocated slightly more memory arrays. In contrast, for attention calculations, which consume data immediately after computation, more compute arrays are allocated. Moreover, after calculating the $K$ value, some $K$ data is stored on-chip as it owns some CIM array in memory mode. Once the respective CIM arrays switch from memory to compute mode, $QK^T$ computations can proceed directly in place, aligning to minimize data transfer. This strategic allocation demonstrates CMSwitch's ability to optimize resource usage and improve performance effectively.

\subsection{Effectiveness for Various Workload Scale}
In this section, we analyze the impact of workload scale on the compilation results of CMSwitch.
\update{As depicted in \fig \ref{fig:scale}, we evaluate BERT-large, Llama 2-7B, OPT-6.7B, and OPT-13B with batch sizes ranging from 4 to 16 and input/output sequence lengths from 32 to 2048.
Meanwhile, we visualize the memory/compute allocation sensitivity across different sequence lengths in \fig \ref{fig:scale} last row. 
The average proportion of arrays operating in memory mode across all segments serves as an indicator of the overall resource allocation strategy, though variations may occur within individual segments.}


\update{As shown in \fig \ref{fig:scale}, across various batch sizes, CMSwitch achieves an average performance improvement of 1.19$\times$ to 1.03 $\times$ over CIM-MLC for BERT when sequence lengths range from 32 to 256. For sequence lengths exceeding 512, CMSwitch demonstrates equivalent performance to CIM-MLC, a trend that remains consistent across different batch sizes. Furthermore, the last row of \fig \ref{fig:scale}  reveals that, as the sequence length increases, the average ratio of arrays in memory mode gradually decreases to zero. This trend aligns with the characteristics of BERT, where the arithmetic intensity increases with longer sequence lengths, necessitating a shift toward compute resources for the corresponding computations. Therefore, as the sequence length extends, CMSwitch’s performance converges with that of CIM-MLC, as we adopt its kernel optimizations.}

\update{For generative models, across various batch sizes, CMSwitch yields average performance gains of 1.25$\times$ to 1.08$\times$ for LLaMA2-7B, 1.50$\times$ to 1.23$\times$ for OPT-6.7B, and 1.76$\times$ to 1.32$\times$ for OPT-13B. However, as sequence lengths extend to 512, the speedup diminishes, with fewer arrays operating in memory mode, after which performance stabilizes. For instance, when evaluating OPT-6.7B with a batch size of 8, and varying input and output sequence lengths from 32 to 2048, the speedups compared to CIM-MLC are 1.41$\times$, 1.58$\times$, 1.27$\times$, 1.23$\times$, 1.22$\times$, and 1.22$\times$, while the corresponding average memory mode array ratios decrease from 20.8\% to 12.0\%. As the input and output sequence lengths increase, the arithmetic intensity during the input processing grows, prompting more arrays to switch to compute mode. This shift reduces the dual-mode switching benefits, resulting in lower overall speedups compared to CIM-MLC, which statically configures all arrays in compute mode.}
\update{To further assess the performance of \name for generative models during different inference stages, we use LLaMA2-7B and OPT-13B as benchmarks. We fix the input length at 128 and vary the output length from 32 to 2048, and vice versa, to observe the speedup. As shown in Figure \ref{fig:scale_vis}, when the input length is fixed, \name achieves a performance improvement of 1.10$\times$ to 1.24$\times$ over CIM-MLC for LLaMA-2 7B and 1.43$\times$ to 1.62$\times$ for OPT-13B, with a nearly consistent speedup as the output sequence length increases.  This consistency arises because the decode phase that generates the output sequence processes tokens incrementally, leaving the arithmetic intensity unaffected by changes in output length. Additionally, the varying memory space required by caching KV with longer output sequences benefits from dynamic mode switching. Conversely, when the output length is fixed, CMSwitch's performance diminishes as the input sequence length increases for both models, driven by the higher arithmetic intensity, which demands additional compute resources.}

\update{Evaluations with varying sequence lengths demonstrate that CMSwitch exhibits adaptability to diverse workload demands by allocating memory and compute resources efficiently, yielding customized compilation results. Moreover, CMSwitch better supports applications with dynamically changing on-chip memory requirements, as existing compilation works typically treat all arrays in compute mode, overlooking their potential use as scratchpads memory.}


%

\subsection{Cost and Scalability Analysis}
\noindent\textbf{Dual-mode switch overhead.}
In our evaluation, the dual-mode switch process introduces negligible overhead, contributing around 3\% - 5\% to the total execution time when providing considerable performance improvement. 
The dual-mode switch process takes time to configure the input drivers for the mode switch. 
This minimal overhead is attributable to the efficient design of the CIM chip, which ensures that the switch between compute and memory modes is both swift and seamless. 
Furthermore, the performance gains realized through our switching overhead-aware network segment strategy, which optimizes network segment scheduling, outweigh the minor switching costs. This evidence confirms that the dual-mode switch mechanism
is a valuable consideration in the CIM compilation optimization process. 

\noindent\textbf{Scalability.}
After adjusting the hardware settings to the PRIME \cite{chi2016prime} architecture, we evaluated the performance of transformer-based networks. Compared to Dynaplasia, PRIME offers larger and more CIM arrays that can contain large network segments. 
However, PRIME has higher write overhead as it uses the ReRAM as the memory device. 
According to our assessment, compared to the CIM-MLC, we achieved speedups of 1.48$\times$ for Bert, 1.09$\times$ for Llama-7B, and 1.10$\times$ for OPT-13B. 
These results indicate that \name can provide performance improvements across different target hardware configurations. This adaptability demonstrates the robustness and efficiency of \name in leveraging the capabilities of various CIM accelerator architectures.

\begin{figure}[t]
    \centering
    \includegraphics[width=0.8\linewidth]{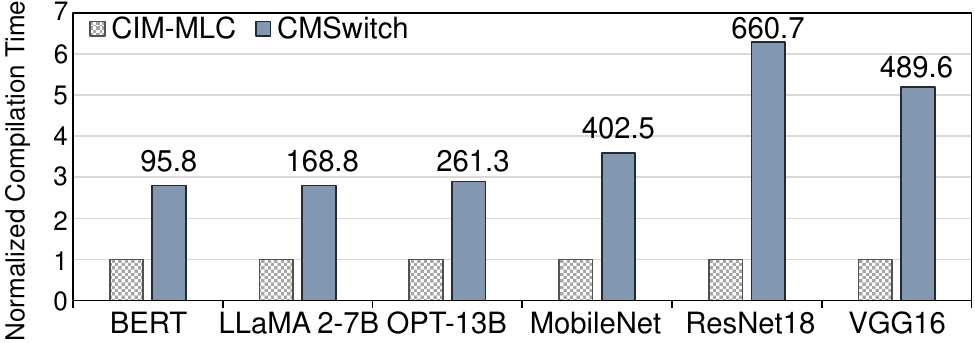}
    \caption{\update{Compilation overhead of different workloads. The number over the bar is the absolute number of compilation time (in s).}}
    \label{fig:overhead}
\end{figure}
\subsection{\update{Compilation overhead}}
\update{We compare the compilation time of CMSwitch with CIM-MLC to demonstrate CMSwitch's overhead. To reduce variability, each benchmark used in the end-to-end evaluation was compiled 20 times. As illustrated in \fig \ref{fig:overhead}, CMSwitch's compilation time is approximately 2.8$\times$ to 6.3$\times$ longer than that of CIM-MLC, indicating a higher compilation overhead. This increase stems from an exponentially expanded compilation space, which introduces more opportunities for optimization. Given that the compilation process is a one-time operation, the extended duration is justified by the potential for substantial performance gains through a more thorough exploration of the optimization space.
Regarding the sensitivity to compilation overhead, we observe that CNNs like ResNet18 and VGG16 require roughly 2.5$\times$ more compilation time compared to transformer-based models. This is attributed to the larger compilation space for CNNs, which feature approximately three times as many convolution types with diverse kernel sizes, whereas transformer-based models allow the compilation results of a single block to be reused across all layers. Consequently, the compilation space for transformers is relatively compact. Therefore, for the evaluated benchmarks, the compilation overhead scales almost linearly with the workload size, even as the optimization space expands. This linear scaling is achieved by leveraging techniques such as impossible-case pruning and non-enumerative solving methods by DP and MIP, which accelerate the exploration of the optimization space and effectively reduce potential overhead.
}

\section{\update{Discussion}}

\textbf{\update{Opportunity for general-purpose system.}}
\update{
While this work focuses on a standalone CIM accelerator system, exploring the compute/memory dual-functionality of CIM architecture offers significant potential for improving general-purpose system performance. Whether CIM is used as a co-processor or a standalone accelerator, dedicating all of its resources solely to computation may not achieve the optimal operating point for many tasks, due to their varying resource requirements.
In a real system enhanced with CIM, tasks that go beyond DNNs will be encountered more frequently \cite{fujiki2019duality,lockerman2020livia,orenes2023dalorex}. 
These tasks often involve complex and dynamic memory and computation needs. To address this, it is crucial to adapt the CIM architecture to meet the specific demands of each task. One effective approach lies in dynamically adjusting the allocation of computation capacity, memory size, and bandwidth through mode switching. This flexibility allows the system to better match the resource requirements of diverse workloads.
The trend toward tighter integration of memory and computation is key to supporting the growing diversity of memory-intensive tasks. 
Therefore, leveraging the dual-mode capability of CIM in a more flexible manner is crucial to optimizing system efficiency.
}


\section{Conclusion}
In this paper, we presented a novel compilation optimization process tailored for dual-mode CIM chips. By incorporating the compute-memory dual-mode switch into the compilation optimization space, we developed a comprehensive approach that optimizes the memory/compute CIM array allocation for various networks. Our dynamic programming-based network segmentation, coupled with mixed integer programming for operator mapping and scheduling, ensures efficient utilization of CIM resources. \update{Experimental results demonstrate 1.31 $\times$ performance improvements on average across different models compared to the state-of-the-art CIM compilation work}, highlighting the effectiveness of our approach in diverse workloads. By utilizing the dual-mode feature of the CIM processor, our method \name achieves substantial speedup while maintaining low overhead. This work paves the way for more efficient and flexible use of CIM chips in real-world DNN applications, enhancing the performance and scalability of CIM-based systems.

\section*{Acknowledgments}
We sincerely thank the anonymous reviewers for their insightful suggestions for improving this paper. This work was partially supported by the National Key R\&D Program of China (Grant No. 2023YFB4404400) and the National Natural Science Foundation of China (Grant No. 62222411, 62025404, 62204164). Ying Wang (wangying2009@ict.ac.cn) is the corresponding author.

\bibliographystyle{plain}
\balance
\bibliography{references}

\end{document}